\documentclass{iaus}
\usepackage{graphicx}

\title[PCA Tomography] 
{PCA Tomography and its application \\ to nearby galactic nuclei
}

\author[J.E. Steiner et al.]   
{J.E. Steiner$^{1\dagger}$, R.B. Menezes$^1$, T.V. Ricci$^1$ \and A.S. Oliveira$^2$}

\affiliation{$^1$IAG, Universidade de S\~ao Paulo, \\
Rua do Mat\~ao, 1226, S\~ao Paulo - SP, Brasil \\[\affilskip]
$^2$IPD/Univap \\ $^{\dagger}${\tt steiner@astro.iag.usp.br}}

\pubyear{2009}
\volume{267}  
\pagerange{119--126}
\jname{Co-Evolution of Central Black Holes and Galaxies}
\editors{B.M.\ Peterson, R.S.\ Somerville, \& T.\ Storchi-Bergmann, eds.}
\begin{document}

\maketitle

\begin{abstract}

With the development of modern technologies such as IFU´s, it is possible to obtain data cubes in which one produces images with spectral resolution. To extract information from them can be quite complex, and hence the development of new methods of data analysis is desirable. We briefly describe a method of analysis of data cubes (data from single field observations, containing two spatial and one spectral dimension) that uses Principal Component Analysis (PCA) to express the data in the form of reduced dimensionality, facilitating efficient information extraction from very large data sets. We applied the method, for illustration purpose, to the central region of the low ionization nuclear emission region (LINER) galaxy NGC 4736, and demonstrate that it has a type 1 active nucleus, not known before. Furthermore, we show that it is displaced from the centre of its stellar bulge. 
\end{abstract}

\firstsection 
\section{Introduction}

With the advent of panoramic spectroscopic devices such as Integral Field Units, it is possible to construct data cubes of immense proportions that present data in three dimensions: two spatial and one spectral. The analysis of these data may become complex and overwhelming, as it may involve tens of millions of pixels. More concerning is that, given this complexity, only some restricted subset of the data ends up being analyzed; the rest is at the risk of being largely ignored. New techniques that allow us to extract information in a condensed, fast and optimized form are therefore necessary and welcome. Here we briefly present a method of data cube analysis that uses Principal Component Analysis (PCA). This method condenses the significant information content associated with the data, through effective dimensional reduction, facilitating its interpretation. PCA compresses the data expressed as a large set of correlated variables in a small but optimal set of uncorrelated variables, ordered by their principal components. Clearly, our shared goal of analyzing data is to extract physical information from them; a dimensional reduction does not necessarily produce valuable information, but an appropriate choice of coordinates may help. PCA is a nonparametric analysis, therefore there are no parameters or coefficients to adjust that somehow depend on the users' experience and skills, or on physical and geometrical parameters of a proposed model. PCA provides a unique and objective answer. PCA has been used many times in the astronomical literature and a more extended presentation of this technique is given in Murtag \& Heck (1987) and Fukunaga (1990). Most of the applications of PCA in astronomy are related to find eigenvectors across a population of objects. In the present case, we want to apply the technique to a single data cube in which the objects are spatial pixels of an individual field, containing a single galaxy, nebula or a set of stars. We identify eigenvectors (the uncorrelated variables) that we refer to as eigenspectra, and tomograms, which are images of the data projected in the space of the eigenvectors. In traditional tomographic techniques, one obtains images that represent 'slices' in tri-dimensional space (the human body, for example) or in velocity space (Doppler Tomography). In PCA Tomography, one obtains images that represent 'slices' of the data in the eigenvectors space (tomograms). Each tomogram is associated with it an eigenspectrum. The simultaneous analysis of both brings a new perspective to the interpretation of them. For a full presentation of PCA Tomography, see Steiner et al (2009).

\section{Eigenspectra and tomograms}

Our aim is to analyze data cubes in which we have two spatial and one spectral dimension. Each pixel of this original three-dimensional data cube has intensity $I_{ij\lambda}$; here i and j define a spatial pixel and $\lambda$ a spectral pixel. Now we organize the data cube $I_{ij\lambda}$ (which has zero mean) into a matrix $I_{\beta \lambda}$ of n rows (spatial pixels, referred to here as objects) and m columns (spectral pixels, referred to here as properties). Then $\beta$ can be expressed as

\begin{equation}
\beta = \mu (i-1) + j
\label{eq1}
\end{equation}

The data cube transformed into the matrix $I_{\beta \lambda}$ will be the subject of the PCA which is a technique used to analyze multidimensional data sets and is quite efficient to extract information from a large set of data as it allows us to identify patterns and correlations in the data that in other ways would hardly be noted. Mathematically, it is defined as a linear orthogonal transformation that expresses the data in a new (uncorrelated) coordinate system such that the first of these new coordinates, $E_1$ (eigenvector 1), contains the largest variance fraction, the second variable, $E_2$, contains the second largest variance fraction and so on. These new coordinates are orthogonal to one another. 

The covariance matrix of $I_{\beta \lambda}$, $C_{Cov}$, is square and has m rows and columns (equal to the number of the original spectral pixels). The main diagonal elements correspond to the variances of each of the isolated variables, while the other (cross) elements correspond to the covariance between two distinct properties. The $m \times m$ covariance matrix has m eigenvectors, $E_k$, each one associated with one eigenvalue, $\Lambda_k$. $E_k$ are the new uncorrelated coordinates and k is the order of the eigenvector that can vary from 1 to m; the eigenvectors are ordered by decreasing value of each associated $\Lambda_k$, which is the variance of each component, to form the characteristic matrix, $E_k$, in which columns correspond to eigenvectors. The transformation that corresponds to the PCA can be represented by the following formula:

\begin{equation}
T_{\beta k} = I_{\beta \lambda} . E_{\lambda k}
\label{eq2}
\end{equation}
where $T_{\beta k}$ is the matrix containing the data in the new coordinate system. As the aim of PCA is to express the original data on the new system of uncorrelated coordinates, one concludes that the ideal covariance matrix of the data in this new coordinate system ($D_{Cov}$)
must be diagonal, that is, the covariance between the coordinates must be zero. One may say that the PCA execution consists in determining the matrix $E_{\lambda k}$ that satisfies equation \ref{eq2} and so that $D_{Cov}$ is diagonal:

\begin{equation}
D_{Cov} = \frac{[T_{\beta k}]^T . T_{\beta k}}{(n-1)}
\label{eq3}
\end{equation}

The diagonal elements of $D_{Cov}$ are the eigenvalues. In the case of data cubes of astronomical interest, it is usual to have two-dimensional images with spectra associated with each spatial pixel. In calculating the PCA of such data cubes, one obtains eigenvectors as a function of wavelength, energy or frequency (properties), that we will also refer to as eigenspectra. On the other hand, $T_{\beta k}$ represents data in a new coordinate system. As our objects are spatial pixels, their projection on to a given eigenvector may be represented as a spatial image. Each column of $T_{\beta k}$ can now be transformed into a two-dimensional image, $T_{ij,k}$ using equation \ref{eq1}. We will refer to these images, $T_{ij,k}$ as tomograms, since they represent `slices' of the data in the space of the eigenvectors. Analyzing tomograms simultaneously with eigenspectra brings together a wealth of information. Spectral characteristics may be identified with features in the image and vice versa. Interpreting such associations facilitates the understanding of the three-dimensional structure within the data cube. 

\section{Application: The central region of the LINER galaxy NGC4736}

Let us illustrate the application of the method to answer the following question: is there a supermassive black hole in the nearby LINER galaxy NGC 4736? LINERs are a class of objects with diverse nature. Although most of them seem to host an AGN in the sense that they are powered by accretion on to a supermassive black hole, some objects have not shown any evidence of this. NGC 4736 is somewhat peculiar because it presents a stellar population that corresponds to an ageing starburst. Could this explain its LINER nature (Cid Fernandes et al., 2004)? We observed this galaxy with the GMOS (Allington-Smith et al. 2002), operated in the IFU mode on Gemini North Telescope. The data cube was obtained using 500 fibres on the object. The spectral resolution was R=2900 covering from 4700 to 6800 $\dot{A}$. Three 20 min integrations were obtained. 

\begin{table}[h]
  \label{tabela1}
  \begin{center}
 {\scriptsize
  \begin{tabular}{ccc}\hline
  Eigenvector & Eigenvalue & Accumulated Fraction \\
  $E_k$ & (\% of the variance) &  (\% of the variance) \\ \hline
  $E_1$ & 99.7443 & 99.7443 \\
  $E_2$ & 0.0883 & 99.8326 \\
  $E_3$ & 0.0325 & 99.8651 \\
  $E_4$ & 0.0129 & 99.8781 \\
  $E_5$ & 0.0084 & 99.8864 \\ \hline
  
  \end{tabular}
  }
    \caption{Eigenvalues of the principal components of NGC 4736.}
 \end{center}
\end{table}

\begin{figure}[h]
\begin{center}
 \includegraphics[width=5.0in]{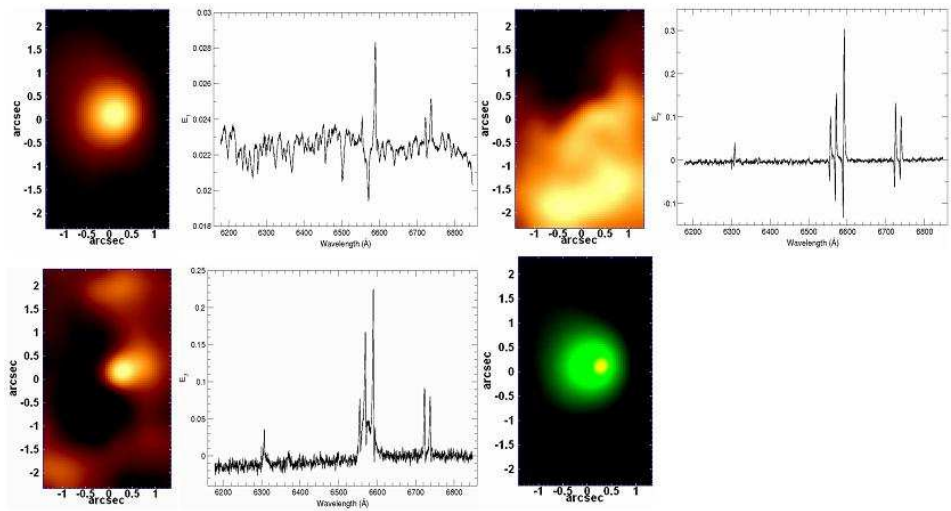} 
 \caption{Tomograms and eigenvectors 1 (upper left, 2 (upper right)and 3 (lower letf). Tomogram 3 (AGN)is superposed on tomogram 1 (bulge) for NGC 4736 showing that the AGN is displaced from the bulge center (lower left).}
   \label{fig1}
\end{center}
\end{figure}

The three principal components are shown as eigenvectors and tomograms in Figure \ref{fig1} and their eigenvalues in Table 1. As can be seen, eigenvector 1 contributes 99.74 per cent of the variance. This means that this eigenspectrum basically replicates what one would see in a spectrum obtained with traditional spectroscopic techniques. Tomogram 1 is the image comparable with that of a classic central stellar bulge. Eigenvector 2 contributes 0.088 per cent of the variance and displays, in combination with its tomogram, a clear map of the rotation of the emission-line gas in the FoV, uncorrelated with the stellar component. Eigenvector 3 contributes 0.032 per cent of the variance. Its characteristic is that it displays correlations among features that can be associated with emission-line transitions. It is quite surprising that features related to two kinds of emission lines are visible: narrow lines, associated with the [OI], [N II] and [SII] species and, also, $H\alpha$. But there is also a feature associated with a broad $H\alpha$ component. This component is typical of Seyfert 1 (or LINER type 1) galaxies and is usually taken as clear evidence for an AGN associated with a supermassive black hole. This has not been reported before, despite the fact that this is a nearby galaxy. 

\section{Discussion and conclusions}

 (i) PCA Tomography identifies eigenvectors, ordered in the form of principal components according to the rank of the corresponding eigenvalues. Tomograms are images that represent 'slices' of the data in the eigenvectors space. The association of tomogram with eigenvectors is important for the interpretation of both. (ii) One of the main advantages of PCA Tomography is the dimensional reduction. This is also important for the data compression and transmission. (iii) The fact that the eigenvectors are orthogonal among themselves is important for their handling and interpretation. When the data cube presents uncorrelated physical phenomena, the orthogonality may be useful for identifying them. (iv) The reconstruction of the data cube with original format, but with separated components associated with distinct eigenvectors, allows the extraction of spectra or images in order to isolate a given feature. (v) Besides, by selecting the eigenvectors or tomograms with certain correlations or anti-correlations, one can enhance features by reconstructing data cubes in original format. (vi) Various types of noise may be eliminated or corrected by selecting their eigenvectors and tomograms: cosmic rays, hot/cold pixels, etc. (vii) We applied it to the central region of the LINER galaxy NGC 4736. The dimensional reduction of the data allowed the identification of characteristics that were unknown in advance. For example, we identify a type 1 nucleus, of very low luminosity, displaced from the centre of the stellar bulge.


\begin{thebibliography}{}

\bibitem[Allington-Smith et al.(2002)]{2002PASP..114..892A} 
Allington-Smith, J., et al.\ 2002, PASP, 114, 892 
\bibitem[Cid Fernandes et al.(2004)]{2004ApJ...605..105C} Cid Fernandes, 
R., et al.\ 2004, ApJ, 605, 105 
\bibitem[Eracleous et al.(2002)]{2002ApJ...565..108E} Eracleous, M., 
Shields, J.~C., Chartas, G., \& Moran, E.~C.\ 2002, ApJ, 565, 108 
\bibitem[Fukunaga (1990)]{book..fuku} Fukunaga K., 1990, Statistical Pattern Recognition, 2nd edn. Academic Press, New York
\bibitem[Heckman (1980)]{1980..AA..87..152} Heckman T. M., 1980, A\&A, 87, 152
\bibitem[Murtag F. \& Heck A.(1987)]{book..murt} Murtag F., Heck A., 1987, Multivariate Data Analysis. Reidel Pub. Co. Dordrecht, Holland
\bibitem[Steiner, J.E. et al.(2009)]{2009MNRAS..395..64} Steiner, J.E., Menezes, R.B., Ricci, T.V. \& Oliveira, A.S. 2009 MNRAS 395, 64

\end{thebibliography}
\end{document}